\documentclass[a4paper]{jpconf}
\usepackage{graphicx}
\usepackage{subfigure}

\begin{document}
\title{Precision analysis of Geant4 condensed transport effects on energy deposition in detectors}

\author{M. Bati\v{c}$^{1,2}$, G. Hoff$^{1,3}$, M. G. Pia$^{1}$}

\address {$^1$ INFN Sezione di Genova, Genova, Italy}
\address{$^2$ Jo\v{z}ef Stefan Institute, Ljubljana, Slovenia}
\address{$^3$ Pontif$\acute{i}$cia Universidade Cat$\acute{o}$lica do Rio Grande do Sul, Brazil}

\ead{Maria.Grazia.Pia@cern.ch}

\begin{abstract}
A comprehensive analysis of the effects of Geant4 algorithms for condensed
transport in detectors is in progress.
The first phase of the project focuses on electron multiple scattering, and
studies two related observables: the longitudinal pattern of energy deposition
in various materials, and the fraction of backscattered particles.
The quality of the simulation is evaluated through comparison with high precision 
experimental measurements; several versions of Geant4 are analyzed to provide
an extensive overview of the evolution of Geant4 multiple scattering algorithms 
and of their contribution to simulation accuracy.
\end{abstract} 

\section{Introduction}

Physics models and computational algorithms operating in the condensed transport
scheme - multiple scattering and energy loss of charged particles - play a
critical role in the simulation of energy deposition in detectors.
Previous studies \cite{tns_sandia,tns_bragg} have highlighted the contribution
of Geant4 \cite{g4nim,g4tns} multiple scattering implementation to simulate the
longitudinal profile of the energy deposited by electrons and protons in matter,
and its effects on the accuracy of the simulation with respect to experimental
measurements.
Nevertheless, despite the relevance of Geant4 condensed transport to
experimental applications, a comprehensive overview of the problem domain is
still missing: quantitative comparisons with experiment for model validation are
relatively scarce in the literature, and the interplay of algorithms and model
parameters on different observables has not been quantified yet.

A project in progress analyzes the effects of algorithms for condensed transport
of electrons in Geant4, namely of multiple scattering.
It evaluates the contribution to simulation accuracy of a variety of Geant4
multiple scattering models with respect to high precision experimental
measurements of the longitudinal pattern of energy deposition, and of the
fraction of backscattered particles.

Geant4 multiple scattering simulation involves a number of parameters and
physics modeling approaches, which have been subject to modifications in the
course of the years.
The results reported in the above mentioned references show significant
variations in the simulation outcome depending on Geant4 versions, which reflect
the ongoing evolution of Geant4.
Due to the fast pace of new Geant4 version releases and correction patches,
several Geant4 versions are concurrently in use in the experimental community;
therefore, the validation analysis of the effects of multiple scattering
implementations is extended to a number of recent Geant4 versions.

This paper reflects the limited amount of information that could be presented in
a poster at the CHEP (Computing in High Energy Physics) 2012 conference; the
full set of results will be extensively documented in a dedicated journal
publication.

\section{Overview of Geant4 multiple scattering models}

Geant4 original multiple scattering model \cite{rd44_1997}, also known as the
``Urban'' model, was conceived as an improvement over the treatment of multiple
scattering in GEANT 3 \cite{geant3}, which was based on Moli\`ere \cite{moliere}
theory.
It belongs to the category of condensed simulation algorithms, which account for
the global effects of a large number of collisions over a track segment,
including the change of direction, net displacement and energy loss of a charged
particle. Geant4 ``Urban'' model is based on Lewis \cite{lewis} theory; details
about its theoretical approach and its original implementation in Geant4 can be
found in \cite{urban_2002}.

A quantitative experimental validation of an early implementation of the
``Urban'' algorithm, concerning high energy muons, is documented in
\cite{arce_2000}: it demonstrates that the algorithm implemented in Geant4 0.1,
with further improvements released in Geant4 1.0, provided better simulation
accuracy than the GEANT 3.15 algorithm.

Geant4 simulation of multiple scattering based on Lewis theory has evolved since
then.
Various ``Urban'' models have been implemented; the latest Geant4 version
(9.5p01) at the time of writing this paper includes four variants of this model,
implemented in the \textit{G4UrbanMscModel90}, \textit{G4UrbanMscModel92},
\textit{G4UrbanMscModel93} and \textit{G4UrbanMscModel95} classes.
Modifications to the original algorithm and results of their application 
are documented in a dedicated report \cite{urban_2006} and in a series of
conference papers 
\cite{em_nss2004}-\cite{radecs_2011}.

The evolution of the ``Urban'' multiple scattering models has affected the
longitudinal pattern of energy deposition: reference \cite{tns_sandia} documents
the simulation accuracy achieved using the electron multiple scattering
algorithm in Geant4 8.1p02 and 9.1 versions, while reference \cite{tns_bragg}
highlights significant differences in the energy deposited in water by protons
of approximately 70 MeV energy over several Geant4 versions from 8.1 to 9.3.
In both these references other Geant4 physics models contributing to determine
the energy deposition patterns were verified not to have changed over the Geant4
versions subject to test, or their evolutions resulted in statistically
insignificant effects; therefore, the observed differences across different
Geant4 versions are likely related to evolutions in the implementation of
multiple scattering.

Additional multiple scattering models have been introduced in later versions of
Geant4 besides the ``Urban'' model: a variant \cite{mscatt_kadri} of the
Gousdmit-Saunderson \cite{goudsmit} model, which according to
\cite{mscatt_chep2009} is intended to provide higher accuracy for electrons and
positrons, and the ``Wentzel-VI'' model, that \cite{mscatt_chep2009}
characterizes as designed for precise simulation of muons and hadrons.

\section{Validation method}

A simultaneous validation is performed to evaluate the accuracy of
backscattering and energy deposition in the same experimental set-up: accurate
rendering of both observables through the same physics settings is a known issue
in Monte Carlo simulation, and a sensitive test of the capabilities of multiple
scattering algorithms.

The validation test exploits two sets of high precision experimental data, taken
by the same experimental group at Sandia laboratories.
The data reported in \cite{sandia79} concern the longitudinal energy deposition
pattern in various materials originating from a beam of electrons with energy
ranging from 0.058~MeV to 1.033~MeV.
The measurements involved beryllium, carbon, aluminium, iron, copper,
molybdenum, tantalum and uranium as targets.
The same experimental set-up was exploited to estimate the fraction of
backscattered energy of the electron beam, reported in \cite{sandia80}: this
quantity was derived from the direct measurement of the total energy deposited
in the target, corrected by the calculated energy escaped from the target volume
as Bremsstrahlung photons.
The experimental configuration for backscattering studies involved beryllium,
carbon, titanium, molybdenum, tantalum and uranium targets.
Further details about the experimental set-up and the characteristics of the
measurements can be found in \cite{sandia79} and \cite{sandia80}.

The simulation configuration reproduced the experimental settings; it is
described in detail in \cite{tns_sandia}.
The energy deposited in the sensitive layers of the target was scored to
determine the energy deposition profile as a function of the electron
penetration depth.
The backscattering fraction $BSF$ was estimated as described in 
\cite{sandia80}:
\begin{equation}
BSF=1 -\frac{\left (D+E_{\gamma}  \right )}{E_{0}},
\label{equation1}
\end{equation}
where $D$ is the total energy deposited in the detector, $E_{\gamma}$ is the
energy escaping from the detector in the form of Bremsstrahlung photons
and $E_{0}$ is the kinetic energy of the incident electrons.

The simulation was configured with a variety of options of electron and photon
interactions available in Geant4: ``Standard'' (identified as \textit{std} in
the figures), based on the evaluated data libraries
distributed by the Lawrence Livermore National Laboratory (identified as
\textit{liv} 
and ``Penelope-like'' (identified as \textit{pen}).
The comparison of the results deriving from the various physics configurations
with the experimental data allows the evaluation of the relative accuracy of
Geant4 electromagnetic modeling options.

Simulated data were produced with Geant4 8.1.p02, 9.1.p03, 9.2.p04, 9.3p02,
9.4.p03 and 9.5 versions.
The longitudinal energy deposition profiles produced by the simulation models in
the low energy electromagnetic package and the multiple scattering
implementation of Geant4 versions 8.1p02 and 9.1 were analyzed in
\cite{tns_sandia} with respect to the experimental data of \cite{sandia79}.
The study in progress extends the analysis to the following Geant4 versions,
also involving the models in Geant4 standard electromagnetic package and new
multiple scattering models developed since the publication of \cite{tns_sandia};
moreover, it complements the validation of the energy deposition profiles with
the analysis of the backscattering fraction.

The accuracy of the simulation is quantitatively estimated by means of a
statistical analysis articulated over two stages: the first one consists of
goodness-of-fit tests to evaluate the compatibility of the simulation results
with experimental data in each test case (experimental configuration, Geant4
physics configuration and version); the second one consists of a categorical
analysis, exploiting contingency tables, to quantify whether the accuracy of
different physics modeling options, or of the same modeling option in different
Geant4 versions, exhibit statistically significant differences.
The analysis method is described extensively in  \cite{tns_sandia}.

%

\section{A sample of results}

The conference poster had room only for a limited sample of representative
results; this paper summarizes the preliminary overview of the recent evolution
and current status of Geant4 that could be presented at the conference.
The complete set of results, also including the quantitative outcome of the
statistical analysis, is meant to be published in a regular peer reviewed
journal.

\begin{figure}
\centering
 \includegraphics[width=10.5cm]{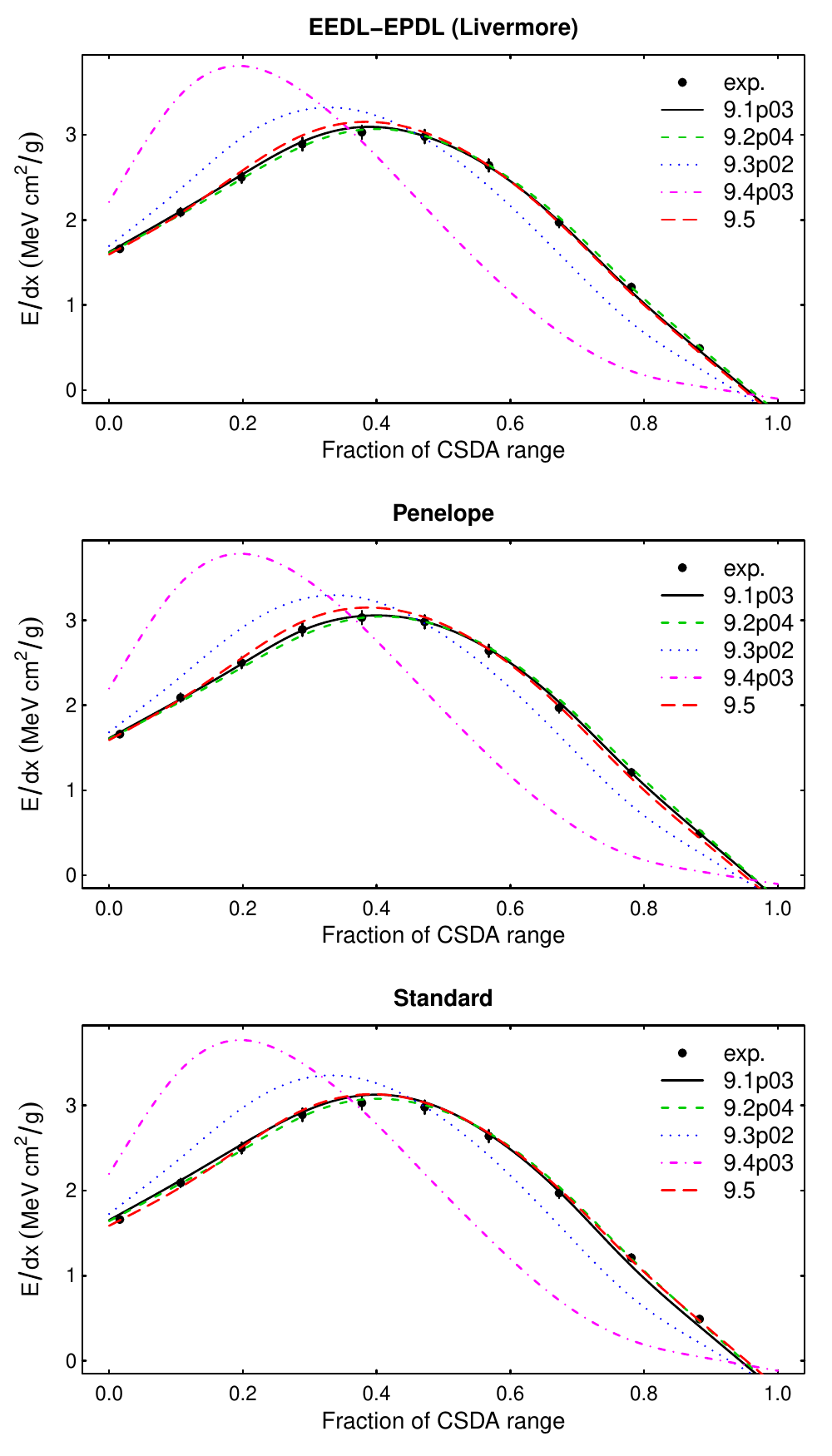}
\caption{Longitudinal profile of energy deposition in 
   carbon for 1~MeV  electrons
   and different Geant4 versions, compared to experimental data.
   The results produced by  various Geant4 modeling options are shown:
``Standard'', based on the EEDL-EPDL evaluated data libraries
distributed by the Lawrence Livermore National Laboratory and
``Penelope-like''. The results of version 8.1 and 9.1, subject to
validation in a previous work \protect{\cite{tns_sandia}}, are included in the
plots to highlight the evolution in Geant4.}
\label{fig_Fig1}
\end{figure}


Figure~\ref{fig_Fig1} shows the absorbed energy profile of 1~MeV electrons in
carbon, produced by different Geant4 versions.
All the simulations results were produced using the default ``Urban'' multiple
scattering model for the corresponding Geant4 version.
The experimental data and the result of the simulation based on Geant4 9.1
documented in \cite{tns_sandia} are reported in all the plots.
Significant discrepancies between simulation and measurements are visible
in the plots concerning Geant4 9.3 and 9.4 versions.

\begin{figure}[h!]
 \centering
 \includegraphics[width=12cm]{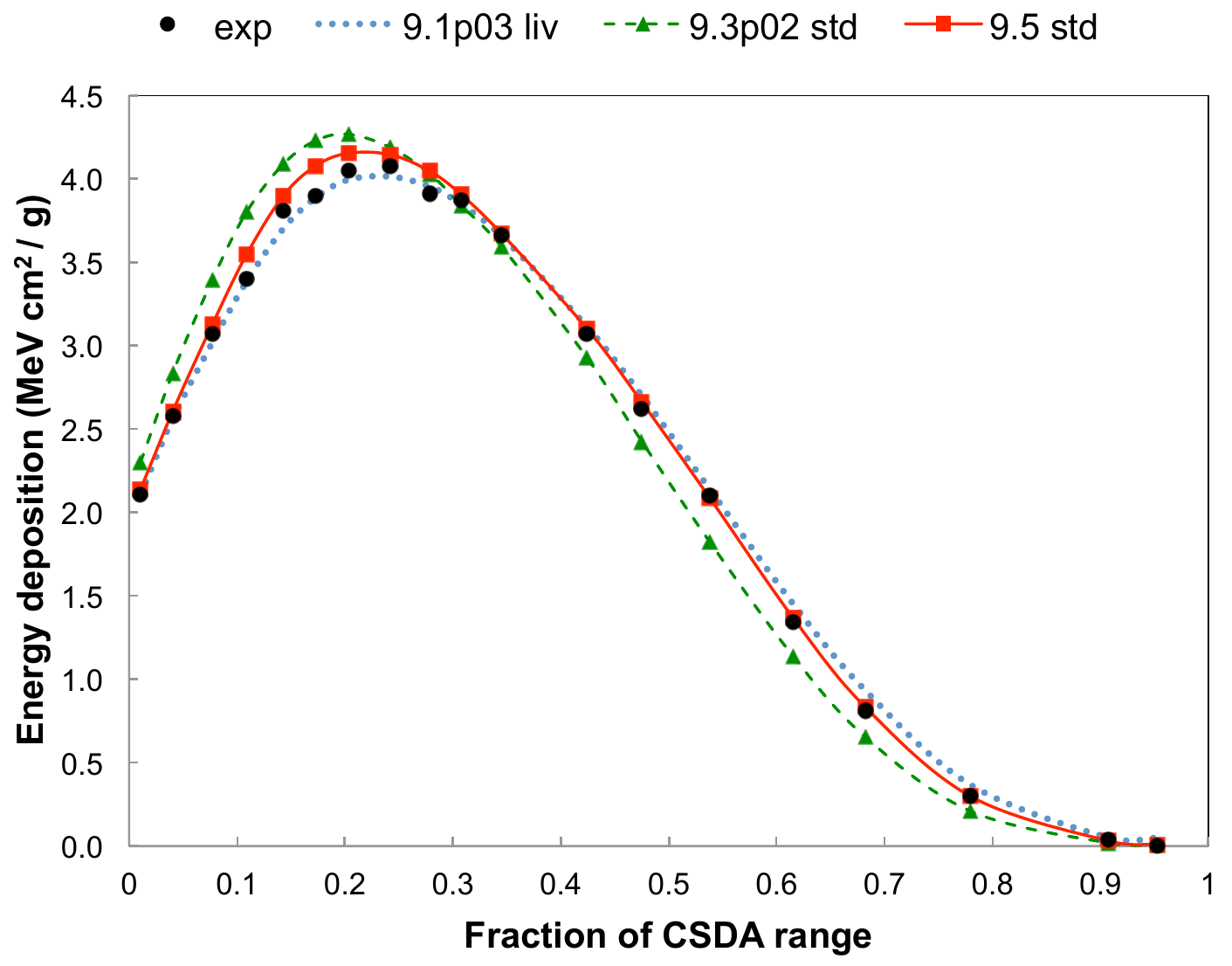}
\caption{Energy deposition of 0.521~MeV electrons in Al: simulation with Geant4
default ``Urban'' multiple scattering algorithm in Geant4 9.5 (red squares) and
Geant4 9.2p03 (green triangles), and experimental data (black circles)
\protect{\cite{sandia79}}. 
The simulation of electron-photon interactions is based on the models in the 
standard electromagnetic package (identified as ``std'') of Geant4 9.2p03 and 9.5.
The blue dotted curve represents the energy deposition profile resulting from
the electromagnetic processes based on the Livermore evaluated data libraries
(identified as ``liv'') and the ``Urban'' multiple scattering model as in Geant4
9.1p03 version.}
\label{fig_almsc}
\end{figure}

Figure \ref{fig_almsc} shows the absorbed energy profile of 521~keV electrons in
aluminium produced with Geant4 9.3p02 and 9.5 versions, activating the default
``Urban'' multiple scattering model and electromagnetic models of the standard
package; the simulated profiles are compared to experimental data.
The plot also shows the profile resulting from the electromagnetic processes
based on the Livermore evaluated data libraries and the ``Urban'' multiple
scattering model as in Geant4 9.1p03 version.
The p-value from the $\chi^2$ test of compatibility with experiment is
1.1$\cdot$10$^{-90}$ for the profile produced by Geant4 9.3p02 and
0.0009 for the profile generated by Geant4 9.5.
The simulation produced with Geant4 9.1p03 version is compatible with experiment
with 0.01 significance.

\begin{figure}[h!]
 \centering
 \includegraphics[width=12cm]{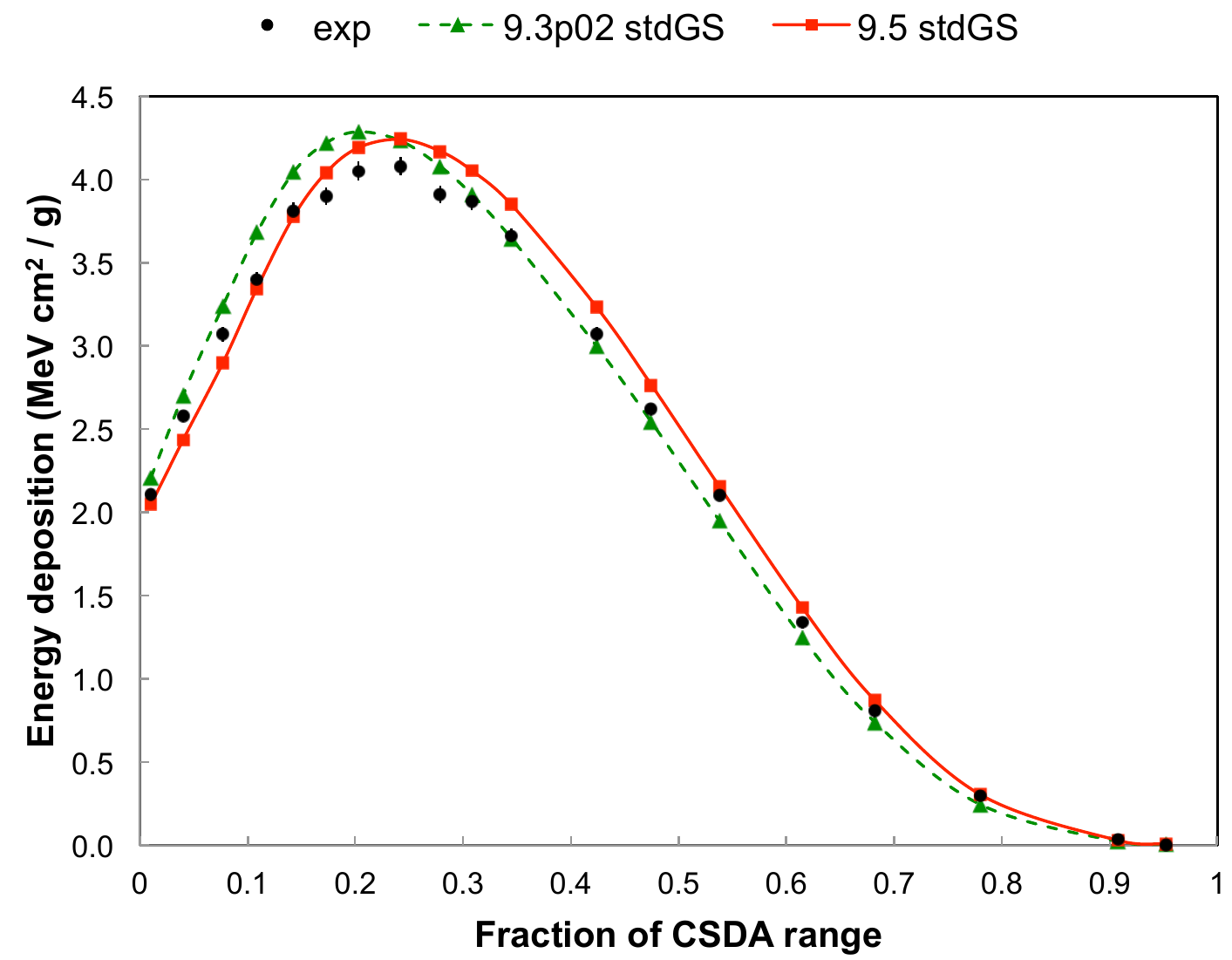}
\caption{Energy deposition of 0.521~MeV electrons in Al, resulting from
simulations with the Goudsmit-Saunderson multiple scattering algorithm in Geant4
9.5 (red squares) and Geant4 9.3p02 (green triangles), and experimental data
(black circles) \protect{\cite{sandia79}}. 
Electron-photon interactions are simulated with Geant4 standard electromagnetic
models (identified as``std'' in the plot). }
\label{fig_al_goudsmit}
\end{figure}

Figure~\ref{fig_al_goudsmit} highlights the evolution of the energy deposition
resulting from simulations where multiple scattering is modeled according to the
Goudsmit-Saunderson algorithm described in \cite{mscatt_kadri}.
The plot shows the energy deposition profiles produced by Geant4 9.3p02 and 9.5,
along with experimental data.
The same test case is illustrated in Figure 5 of reference \cite{mscatt_kadri},
whose publication predates the release of Geant4 9.3.
The simulation results in Figure \ref{fig_al_goudsmit} appear qualitatively
different from those shown in the article documenting the original
implementation of the Goudsmit-Saunderson multiple scattering model
\cite{mscatt_kadri}.
The p-value from the $\chi^2$ test of compatibility with experiment is
1.6$\cdot$10$^{-34}$ for the profile produced by Geant4 9.3p02 and
1.8$\cdot$10$^{-22}$ for the profile generated by Geant4 9.5.

Table \ref{tabone} reports preliminary results of the backscattering fraction,
estimated according to equation \ref{equation1}, for a few target materials and
electron beam energies; the simulation was produced with Geant4 9.3p02 version.
Large differences are visible with respect to the results of \cite{sandia80};
the origin of these discrepancies is under investigation.

\begin{table}[h!]
\centering
\caption{Backscattering fraction.}  
\begin{tabular}{ c c c c }
\hline
Material	&Beam    	&BSF 		&BSF \\
         	&Energy    &Experiment   	&Geant4 9.3p02\cr 
\hline
Tantalum &1 MeV  &0.191 &0.310 \\
Carbon &1 MeV  &0.028 &0.020 \\
Carbon &25 keV   &0.103 &0.296 \\ 
\hline
\label{tabone}
\end{tabular}

\end{table}

The ``efficiency'' of a Geant4 electromagnetic physics model is defined as the
fraction of test cases in which the $\chi^2$ test does not reject the null
hypothesis of compatibility between the energy deposition profile simulated by
that model and experimental measurements, at 0.01 level of significance: it
quantifies the capability of that simulation model to produce results
statistically consistent with experiment over the whole experimental data sample
of \cite{sandia79} involved in the validation process.

The efficiency at reproducing the measured longitudinal energy deposition
profiles of \cite{sandia79} is summarized in Table \ref{tab_eff} for Geant4
9.5p01, which is the latest Geant4 stable version at the time of
submitting this paper to CHEP 2012 proceedings, and for the Geant4 version
validated in \cite{tns_sandia}.
The efficiencies reported in the table correspond to the three Geant4
electromagnetic models (``Livermore'', ``Penelope-like'' and ``Standard'')
associated with default electron multiple scattering settings in the respective
Geant4 version.
This table provides guidance to Geant4 users on which electromagnetic model to use in the latest
version of Geant4 to achieve higher accuracy, and shows how the simulation
accuracy achievable in the latest version has evolved with respect
to the performance in a previously validated Geant4 version.

\begin{table}[h!]
\centering
\caption{Efficiency of Geant4 models at reproducing experimental energy deposition profiles. }  
\begin{tabular}{ l  c c }
\hline
Model	& Geant4 9.5p01 		& Geant4 9.1p03\\
\hline	
Livermore	& 0.47	$\pm$ 0.09	& 0.73	$\pm$ 0.08 \\
Penelope	& 0.13	$\pm$ 0.06 	& 0.30	$\pm$ 0.08 \\
Standard	& 0.17	$\pm$ 0.07	& 0.17	$\pm$ 0.07 \\
\hline
\label{tab_eff}
\end{tabular}

\end{table}

\section*{Conclusions}
The preliminary results summarized here show that different versions of Geant4
produce visibly different energy deposition profiles resulting from the
interaction of electrons with energy up to 1 MeV, when the same conditions
(geometry, beam energy, target material, nominal physics configuration) are
compared.
Statistical analysis based on goodness-of-fit tests confirms that the energy
deposition distributions produced by some Geant4 versions exhibit a significant
disagreement with respect to experimental data.
Incompatibility with the experimental reference is also evident in a preliminary
analysis of the backscattering fraction calculated by one of such versions.

The similar behaviour exhibited by different Geant4 electromagnetic models in
the context of the same Geant4 version suggests that energy deposition patterns
are strongly affected by the evolution of the implementation of the \textit{Urban}
multiple scattering model,
which is common to the simulations activating different electron-photon
interaction models.

Variations in the energy deposition patterns are visible when different 
multiple scattering models are activated (e.g. \textit{Urban} and
\textit{Goudsmit-Saunderson}. This observation suggests
that factors other than multiple scattering algorithms also contribute 
to the observed different behaviour with respect to experimental measurements
resulting from different Geant4 versions.

Further work is in progress to complete the statistical data analysis, that
quantifies the evolution and current status of Geant4 simulation accuracy at
reproducing the experimental data of \cite{sandia79} and \cite{sandia80}.
The results will be documented in a forthcoming publication.

\section*{Acknowledgments}
This work has been partly funded by CNPq BEX6460/10-0 grant, Brazil.

\end{document}